\documentclass[preprintnumbers,aps,prd,preprint,superscriptaddress,nofootinbib]{revtex4-1}

\usepackage{booktabs}
\usepackage{footnote}
\usepackage{lipsum}
\usepackage{amsmath}    
		\usepackage{amssymb}    
\usepackage{color}         
\usepackage{graphicx}     
\usepackage[T1]{fontenc}
\usepackage{times}         
\usepackage{mathrsfs} 
\usepackage{myterms}
\usepackage{hyperref} 
\hypersetup{
colorlinks=true,       
linkcolor=blue,        
citecolor=blue,        
filecolor=magenta,     
urlcolor=blue          
}
\usepackage[all]{hypcap} 

\usepackage[none]{hyphenat} 
\usepackage{ upgreek } 

\usepackage{color,xcolor,fancybox,epsf,rotating,colordvi}

\newcommand{\SLASH}[2]{\makebox[#2ex][l]{$#1$}/}

\newcommand{\pslash}{\SLASH{\vec p}{.2}}

\newcommand{\nmssm}{{\rm NMSSM}}
\newcommand{\mssm}{{\rm MSSM}}
\newcommand{\SMrm}{{\rm SM}}
\newcommand{\exrm}{{\rm ex}}
\newcommand{\Rerm}{{\rm Re}}
\newcommand{\Diagrm}{{\rm Diag}}

\newcommand{\SUtwoL}{{\rm SU(2)_L}}
\newcommand{\UoneY}{{\rm U(1)_Y}}

\newcommand{\dltamu}{\delta a_\mu}
\newcommand{\dltac}{\delta_c}
\newcommand{\dltan}{\delta_n}
\newcommand{\mugmt}{{\rm muon}~ g\!-\!2}
\newcommand{\mmu}{m_\mu}
\newcommand{\tanb}{\tan\!\beta}
\newcommand{\funN}{F^N}
\newcommand{\funC}{F^C}
\newcommand{\sqrtwo}{\sqrt{2}}
\newcommand{\sqrhalf}{\frac{1}{\sqrt{2}}}

\newcommand{\Shat}{\hat{S}}
\newcommand{\Hhat}{\hat{H}}
\newcommand{\Qhat}{\hat{Q}}
\newcommand{\Lhat}{\hat{L}}
\newcommand{\uhat}{\hat{u}}
\newcommand{\dhat}{\hat{d}}
\newcommand{\ehat}{\hat{e}}

\newcommand{\smu}{\tilde{\mu}}
\newcommand{\snu}{\tilde{\nu}}
\newcommand{\cino}{\tilde{\chi}^\pm}
\newcommand{\nino}{\tilde{\chi}^0}
\newcommand{\wino}{\tilde{W}}
\newcommand{\bino}{\tilde{B}}
\newcommand{\hino}{\tilde{H}}
\newcommand{\sino}{\tilde{S}}

\begin{document}

\preprint{\parbox{1.6in}{\noindent arXiv:}}

\title{A smuon in the NMSSM confronted with the muon g-2 and SUSY searches}

\author{Kun Wang}
\email[]{wk2016@whu.edu.cn}
\affiliation{College of Science, University of Shanghai for Science and Technology, Shanghai 200093, China}

\author{Jingya Zhu}
\email[]{zhujy@henu.edu.cn} 
\affiliation{School of Physics and Electronics, Henan University, Kaifeng 475004, China}

\date{\today}

\begin{abstract}
Motivated by the recent supersymmetry (SUSY) search results which prefer most SUSY particles heavy, and muon g-2 anomaly which prefers colorless SUSY particles light, we explore the status of a light smuon (the SUSY partner of left-handed muon lepton) in the Next-to-Minimal Supersymmetric Standard Model (NMSSM). 
Assuming colored SUSY particles be heavy, and considering numerous experimental constraints including muon g-2, SUSY searches, and dark matter, we scan the parameter space in the NMSSM with $\mathbb{Z}_3$-symmetry, checking the status of colorless SUSY particles and their possible mass order, and paying special attention to the smuon. 
After calculation and discussions, we draw the following conclusions:
(i)
	The dominated SUSY contributions to muon g-2 come from the chargino-sneutrino loops, but the muon g-2 anomaly can also constrain the mass of smuon seriously because of the mass-degenerate relation between smuon and muon sneutrino. 
	To interpret the muon g-2 anomaly at $1\sigma$ ($2\sigma$) level, the smuon need to be lighter than $1\TeV$ ($1.8\TeV$). 
(ii)
	When $\nino_1$ is wino- or higgsino-dominated, the smuon can escape from the direct searches with low mass, e.g., $300\GeV$. 
(iii)
	When smuon and $\nino_1$ are mass-degenerate, the smuon can be as light as $200\GeV$, while the $\nino_1$ is usually bino-dominated, or some singlino-dominated, and its relic density can most likely reach the observed value, and the dominated annihilating mechanism is slepton annihilations. 
	In addition, we also list several benchmark points for further studies. 
\end{abstract}


\maketitle

\section{Introduction}
\label{sec:intro}
Recently in April 2021, the new result of muon anomalous magnetic moment ($\mugmt$) at Fermilab was reported \cite{Muong-2:2021ojo}, showing an exciting disparity from the SM prediction. 
Combined with the previous result at Brookhaven \cite{Muong-2:2006rrc}, the discrepancy can be $4.2~\sigma$ \cite{Muong-2:2021ojo}: 
\begin{eqnarray}
	\delta a_\mu \equiv a_\mu^{\exrm} \!-\! a_\mu^{\SMrm} 
	\;= (25.1 \!\pm\! 5.9) \times\! 10^{-10} \;, 
\end{eqnarray}
where the combined experimental result $a_\mu^{\rm ex}$ and the SM prediction $a_\mu^{\rm SM}$ are given by 
\begin{eqnarray}
	a_\mu^{\exrm} &=& (11659206.1 \!\pm\! 4.1) \times\! 10^{-10} \;, \\
	a_\mu^{\SMrm} &=& (11659181.0 \!\pm\! 4.3) \times\! 10^{-10} \;. 
\end{eqnarray}

It is widely believed that the Standard Model (SM) is not a complete description of nature. 
Besides the $\mugmt$ anomaly, SM can also not interpret the hierarchy, grand unification, dark matter (DM), and baryogenesis problems, etc. 
To interpret these problems, theorists proposed many theories beyond the SM (BSM), among which the supersymmetry (SUSY) is a most fascinating one \cite{Haber:1984rc, Martin:1997ns}. 
SUSY assumes that every SM particle has a SUSY partner, whose spin differs from the corresponding SM particle by a half. 
These extra SUSY particles can contribute a lot to solving the problems and anomalies including $\mugmt$. 
The SUSY effects to $\mugmt$ were first calculated in 1980s \cite{Grifols:1982vx, Ellis:1982by, Barbieri:1982aj}, and were comprehensively expressed in Ref. \cite{Moroi:1995yh, Martin:2001st, Stockinger:2006zn}. 
In recent years, and especially after the new result released at Fermilab, there have been numerous papers working on interpreting the $\mugmt$ anomaly in the Minimal Supersymmetric Standard Model (MSSM) and its extensions such as the Next-to MSSM (NMSSM) \cite{Shafi:2021jcg, Li:2021pnt, Cao:2021tuh, Wang:2021bcx, Abdughani:2021pdc, Han:2021ify, Cox:2021nbo, Ahmed:2021htr, Iwamoto:2021aaf, Zheng:2021wnu, Zhao:2021eaa, Su:2020lrv, Wang:2020tap, Wang:2020dtb, Badziak:2019gaf, Wang:2018vxp, Wang:2018vrr, Talia:2017twv, Wang:2017vxj, Badziak:2014kea, Cao:2012fz, Cao:2012yn, Cao:2011sn}. 
It is found that, in MSSM and NMSSM, the main contributions to $\delta a_\mu$ come from the chargino-sneutrino loops and the neutralino-smuon loops. 
Thus generally, light SUSY particles, such as smuons, sneutrinos, charginos, neutralinos, are required to interpret the distinct $\mugmt$ anomaly. 

However, the null results searching for SUSY at the LHC in recent years set strong constraints to SUSY particles. 
For gluino and the first two-generation squarks, the low mass bounds can be over $2\TeV$ \cite{ATLAS:2019fag, Vami:2019slp, ATLAS:2021twp}. 
Even for the colorless charginos and sleptons including smuon, the low mass bounds can be hundreds of GeV in simple models except for compressed-spectrum scenario \cite{ATLAS:2021yqv, CMS:2021cox, CMS:2017moi, ATLAS:2018ojr, ATLAS:2019lff, CMS:2018eqb}. 
In detail, the recent searches for smuon by the ATLAS collaboration with $139\fbm$ data shows that, the mass bound can be about $600\GeV$ assuming smuon decay $100\%$ to muon plus the lightest neutralino \cite{ATLAS:2019lff}. 
Thus for constraints to SUSY models, the direct SUSY searches at the LHC and the $\mugmt$ anomaly may be in tension with each other. 
In this work, we consider the status of smuon in NMSSM confronted with $\mugmt$ and SUSY searches, and their implications for dark matter in addition. 

The rest of this work is reminded as follows. 
In \sref{sec:model} we introduce the model NMSSM, presenting relevant analytic calculations of $\mugmt$ 
briefly. 
In \sref{sec:result} we state the numerical calculations and carry out the discussions. 
Finally, we summary and draw our principal conclusions in \sref{sec:summary}.

\section{The model and analytic calculations}
\label{sec:model}
Instead of a manual and massive $\mu$ parameter in MSSM, in NMSSM the $\mu$ parameter is effective and generated naturally when the additional singlet superfield $\Shat$ gets a vacuum expected value (VEV) \cite{Ellwanger:2009dp, Maniatis:2009re}. 
In the NMSSM of $\mathbb{Z}_3$-invariant vision, the superpotential can be expressed as 
\begin{eqnarray}
	W^\nmssm = W^\mssm_{\mu\to\lambda\Shat} + \frac{\kappa}{3}\Shat^3
\end{eqnarray}
where $\lambda$ and $\kappa$ are dimensionless couplings, $W^\mssm$ is the superpotential in MSSM 
\begin{eqnarray}
	W^\mssm = \mu \Hhat_u \!\!\cdot\!\! \Hhat_d 
	\!+\! {\bf y\!_u} \uhat \Qhat \!\!\cdot\!\! \Hhat_{\!u} 
	\!-\! {\bf y\!_d} \dhat \Qhat \!\!\cdot\!\! \Hhat_{\!d} 
	\!-\! {\bf y\!_e} \ehat \Lhat \!\!\cdot\!\! \Hhat_{\!d}, 
	~~~~~~
\end{eqnarray}
and in terms including quarks or leptons, the generation indexes being summed over are omitted. 

After electroweak symmetry breaking, several SUSY particles with the same quantum numbers mix, generating an equal number of mass eigenstates. 
The SUSY partners of left- and right-handed muon leptons have spin 0, both called smuon. 
In the $\{ \smu_L, \smu_R \}$ basis, the smuon-mass matrix can be written as
\begin{eqnarray}
M^2_{\smu}
\! =\! \left( \begin{array}{cc} \!\!\! 
		M_L^2 \!+\! (\sin^2\!\theta_{W} \!-\! \frac{1}{2}) m_Z^2 \! \cos\!2\beta 
		& m_\mu (A_{\smu}\!-\!\mu\tan\beta) 
		\\
		m_\mu (A_{\smu}\!-\!\mu\tan\beta) 
		& M_R^2 \!-\! m_Z^2\sin^2\!\theta_{W} \! \cos\!2\beta \!\!\!
	\end{array}\right) \!, ~~~~
\end{eqnarray}
where we can see that when $M_{L,R}\gg m_Z, \sqrt{m_\mu A_{\smu}}, \sqrt{m_\mu \mu \tan\!\beta}$, the mixing between $\smu_L$ and $\smu_R$ is negligible. 
Since in NMSSM the neutrinos are left-handed and massless, the mass of muon neutrino's partner muon sneutrino $m_{\tilde{\nu}}$ is related to the left-handed smuon mass by 
\begin{eqnarray}
	m^2_{\snu} &=& M_L^2  +  \frac{1}{2}m^2_Z\cos\!2\beta \;.
\end{eqnarray}
When $m_L\gg m_Z$, the masses are approximately equal. 

The SUSY partners of gauge bosons $W_{1,2,3}, ~B^0$, doublet Higgs, and singlet scalar, are called winos, bino, higgsinos, and singlino, respectively, and all have spin $1/2$ and are written with a tilde. 
Among them, the charged ones mix to two pair of charginos, and the neutral ones mix to form five neutralinos. 
In the base of $\{ \tilde{W}^\pm, \tilde{H}^\pm \}$, the chargino-mass matrix can be written as
\begin{eqnarray}
M_{\cino} 
= \left( \begin{array}{cc} 
	M_2   &   \sqrtwo\,m_W \sin\!\beta \\ 
	\sqrtwo\,m_W \cos\!\beta  &   \mu  
\end{array} \right)
\label{cinomass}
\end{eqnarray}
In the base of $\{ \bino^0,  \wino^0, \hino_u^0, \hino_d^0, \sino \}$, the neutralino-mass matrix can be written as 
\begin{eqnarray}
\label{ninomass}
M_{\nino}
=   \left(  \begin{array}{ccccc}
 M_{1} & 0 & -m_Z \cos\!\beta \sin\!\theta_W  & m_Z \sin\!\beta\sin\!\theta_W & 0 \\
 0 & M_{2} & m_Z \cos\!\beta \cos\!\theta_W & -m_Z \sin\!\beta\cos\!\theta_W & 0 \\
 -m_Z \cos\!\beta\sin\!\theta_W & m_Z \cos\!\beta\cos\!\theta_W & 0 & -\mu & -\lambda v_{d} \\
 m_Z \sin\!\beta\sin\!\theta_W & -m_Z \sin\!\beta\cos\!\theta_W & -\mu & 0 & -\lambda v_{u} \\
 0 & 0 & -\lambda v_d & -\lambda v_u & 2\kappa \mu/\lambda \\
\end{array}  \right)
~~~
\end{eqnarray}

The SUSY effects on $\mugmt$ at the one-loop level are mainly from the chargino-sneutrino loops and the neutralino-smuon loops, with summations performed over all corresponding mass eigenstates. 
The contributions of these two kinds of loops can be writen respectively as \cite{Moroi:1995yh}, 
\begin{eqnarray}
\delta_n & = & \frac{m_\mu}{16\pi^2}  \sum_{i,m} 
\left[ 
	-\frac{m_\mu}{12m^2_{\smu_m}} \left( |n_{im}^L|^2 \!+\! |n^R_{im}|^2\right) F^N_1(x_{im}) 
	+\frac{m_{\nino_i}}{3 m^2_{\smu_m}} \Rerm\left(n^L_{im}n^R_{im}\right) F^N_2(x_{im})
\right]
\label{deltan}
\\
\delta_c & = & \frac{m_\mu}{16\pi^2}\sum_k 
\left[ 
 \frac{m_\mu}{ 12 m^2_{\snu_\mu}} \left(|c^L_k|^2 \!+\! |c^R_k|^2 \right) F^C_1(x_k)
 +\frac{2m_{\cino_k}}{3m^2_{\snu_\mu}} \Rerm \left(c^L_kc^R_k\right) F^C_2(x_k)
\right]
\phantom{deltac}
\end{eqnarray}
where the variables $x_{im}=m_{\nino_i}^2/m_{\smu_m}^2$,  $x_k=m_{\cino_k}^2/m_{\snu_\mu}^2$, with $i=1\!\sim\!5$ denote five mass-eigenstate neutralino labels, and $k=1,2$, $m=1,2$ denote two mass-eigenstate chargino and smuon labels respectively. 
The functions depending on $x_{im}, x_k$ are from loop integrals and  normalized, can be writen as \cite{Moroi:1995yh}
\begin{eqnarray}
	\funN_1(x)&=& \frac{2}{(1-x)^4} \left( 1-6x+3x^2+2x^3-6x^2\ln x \right) \,, ~~~~~~
	\\
	\funN_2(x)&=& \frac{3}{(1-x)^3} \left( 1-x^2+2x\ln x \right) \,,
	\\
	\funC_1(x)&=& \frac{2}{(1-x)^4} \left( 2+ 3x - 6x^2 + x^3 +6x\ln x\right) \,,
	\\
	\funC_2(x)&=&-\frac{3}{2(1-x)^3} \left( 3-4x+x^2 +2\ln x \right) \; . 
\end{eqnarray}
With these functions, $F_{1,2}^{N,C}(1) \!=\! 1$ correspond to the case that SUSY particles degenerate. 
In addition, the corresponding coefficients are writen as 
\begin{eqnarray}
	n^R_{im} &=&  \sqrtwo g_1 N_{i1} X_{m2} + y_\mu N_{i3} X_{m1}
	\\
	n^L_{im} &=&  \sqrhalf \left(g_2 N_{i2} \!+\! g_1 N_{i1} \right) X_{m1}^* - y_\mu N_{i3} X^*_{m2}
	\\
	c^R_k &=& y_\mu U_{k2} 
	\\
	c^L_k &=& -g_2V_{k1}, 
\end{eqnarray}
where $N_{ij}$, $\{U_{kl}, V_{kl}\}$, and $X_{mn}$ are the mixing matrix of neutralino, chargino, and smuon respectively, defined as 
\begin{eqnarray}
	N^* M_{\nino} N^\dagger &=& \Diagrm 
	\left\{ m_{\nino_1}, m_{\nino_2}, m_{\nino_3}, m_{\nino_4} \right\}
	\\
	U^* {M}_{\cino} V^\dagger &=& \Diagrm 
	\left\{ m_{\cino_1}, m_{\cino_2} \right\}
	\\
	X M^2_{\smu}\, X^\dagger &=& \Diagrm 
	\left\{ m^2_{\smu_1}, m^2_{\smu_2} \right\} \;.	
\end{eqnarray}
Finally, $g_2 \!\simeq\! 0.66$ and $g_1 \!\simeq\! 0.36$ are the $\SUtwoL$ and $\UoneY$ gauge couplings respectively, and $y_\mu \ll g_{1,2}$ is the muon Yukawa coupling.   

Considering SUSY particles in the loops, which are neutralinos $\nino_{1,2,3,4,5}$ and smuons $\smu_{1,2}$, or charginos $\cino_{1,2}$ and muon sneutrino $\snu_{\mu}$, approximately have the same masses, $\sqrt{m_{\nino} m_{\smu}}$ or $\sqrt{m_{\cino} m_{\snu_\mu}}$, 
their contributions to $\mugmt$ can be approximately writen as 
\cite{Martin:2001st} 
\begin{eqnarray}
	\label{eq:deltan}
	\delta_n &\simeq& \frac{1}{192\pi^2} \frac{\mmu^2}{m_{\nino} m_{\smu}}(g_1^2 \!-\! g_2^2) \tanb \,,
	\\
	\label{eq:deltac}
	\delta_c &\simeq& \frac{1}{32\pi^2} \frac{\mmu^2}{m_{\cino} m_{\snu_{\mu}}}g_2^2 \tanb \;.
\end{eqnarray}

\section{Numerical calculations and discussions}
\label{sec:result}

In this work, we first scan the parameter space of NMSSM with $\mathbb{Z}_3$-symmetry to get surviving samples under corresponding theoretical and experimental constraints. 
To focus on the light smuon, one of the light colorless SUSY particles preferred by the $\mugmt$ anomaly, we assume colored SUSY particles and heavy Higgs to be heavy, and fix them to $5\TeV$: 
\begin{eqnarray}\label{eq:fix}
	M_A,M_3,M_{Q_i},M_{U_i},M_{D_i} = 5\TeV \,,
	\quad 
	A_{U_i},A_{D_i} = 0 \TeV \,, 
\end{eqnarray}
with the generation index $i=1,2,3$. 
The other parameters correlated with light and colorless SUSY particles we need are set as 
\begin{gather}
	\label{eq:space}
	0<\lambda, |\kappa|<0.8, \quad 1<\tan\beta <100 \,,
	\nonumber\\
	0<\mu, M_1, M_2, M_P<1\TeV \,,
	\nonumber\\
	0<M_{L_i},M_{E_i},|A_{E_i}|<10\TeV \,. 
\end{gather}
To be much different from former studies that $\tilde{\uptau}_1$ as the lightest slepton, we also require manually $m_{\tilde{\uptau}_1} > m_{\smu_1}$. 

The model spectrum including masses and decay information are calculated with {\sf NMSSMTools-5.5.4} \cite{Ellwanger:2004xm, Ellwanger:2005dv, Das:2011dg}. 
The corresponding constraints we consider in our scan can be listed briefly as follows: 
\begin{itemize}
\item Theoretical constraints from a stable vacuum and inexistent Landau pole below the GUT scale \cite{Ellwanger:2004xm, Ellwanger:2005dv, Ellwanger:2006rn}. 
	
\item LEP constraints on charginos and sleptons mass that is heavier than about $100\GeV$. 
	
\item Flavour constraints, such as rare the B-meson decays and the mass differences $\Delta m_d$, $\Delta m_s$ \cite{Tanabashi:2018oca,  Aaij:2012nna, Lees:2012xj, Lees:2012ym}.
	
\item A Higgs boson of $123 \sim 127 \GeV$ with signal predictions consistent globally with Higgs data at LHC \cite{Aad:2019mbh, Sirunyan:2018koj, Khachatryan:2016vau}. 
	
\item Constraints from searches for additional Higgs and exotic decay of the SM-like Higgs, imposed with \textsf{HiggsBounds-5.10.1} \cite{Bechtle:2015pma, Bechtle:2013wla, Bechtle:2013gu, Bechtle:2011sb, Bechtle:2008jh}.
	
\item Upper bounds of DM relic density $\Omega h^2 \le 0.131$ from WMAP/Planck \cite{Hinshaw:2012aka, Ade:2013zuv}, 
	with the relevant quantities calculated by \textsf{micrOMEGAs 5.0}
	\cite{Belanger:2006is, Belanger:2008sj, Belanger:2010pz, Belanger:2013oya}.
	
\item Constraints of direct searches for DM, on the spin-independent case by XENON1T \cite{Aprile:2018dbl}, and spin-dependent case by LUX \cite{Akerib:2016lao}, XENON1T \cite{Aprile:2019dbj} and PICO-60 \cite{Amole:2019fdf}, where the original values are rescaled by  $\Omega/\Omega_0$ accordingly.
	
\item The constraints from $\mugmt$ at $2\sigma$ level, where we use the combined experimental value $a_\mu^{\exrm} = (11659206.1 \!\pm\! 4.1) \!\times\! 10^{-10}$ \cite{Muong-2:2021ojo}, and calculate the SM value $a_\mu^{\SMrm}$ without Higgs contribution, considering the NMSSM contains a SM-like Higgs. 
	Including also the theoretical error of SUSY contributions, we get the range of $\delta a_\mu\equiv a_\mu^{\exrm} -a_\mu^{\SMrm}$ as $(9.5\!\sim\! 41.5) \!\times\! 10^{-10}$ at $2\sigma$ level. 
\end{itemize}

Moreover, in imposing the constraints from SUSY searches, we exploit \textsf{SModelS 2.1.1} \cite{Kraml:2013mwa, Ambrogi:2017neo, Ambrogi:2018ujg, Dutta:2018ioj, Buckley:2013jua, Sjostrand:2006za, Sjostrand:2014zea,  Alguero:2020grj, Khosa:2020zar} with official database version \textsf{2.1.0} \cite{Alguero:2021dig}. 
To focus on the status of smuon, we pay special attention to the constraints of smuon searches in Refs. ATLAS-SUSY-2018-32 \cite{ATLAS:2019lff} and CMS-SUS-17-009 \cite{CMS:2018eqb}. 
In the analyses about smuons, it is assumed to be produced in pairs and $100\%$-ly decay into a muon lepton and the lightest neutralino, thus the final state is composed of two opposite-charge and same-flavor leptons (dimuon) and large missing transverse momentum $\pslash_T$. 
But unfortunately, in both reports, no significant excess over the SM background is observed. 
The ATLAS analysis is based on $139\fbm$ data at the $13\TeV$ LHC, and smuon mass up to about $600\GeV$ are excluded with massless lightest neutralino. 
To consistent with the analyses, we employ \textsf{Resummino 3.0}  \cite{Fiaschi:2019zgh, Fiaschi:2018hgm, Fuks:2013vua, Bozzi:2006fw, Bozzi:2007qr, Fuks:2013lya, Fiaschi:2018xdm} to calculate at next-to leading order (NLO) the smuon production at the 13-TeV LHC: 
\begin{equation}
	p p  \to \smu^+_L \smu^-_L,\qquad p p \to \smu^+_R  \smu^-_R \, , 
\end{equation}
where $\smu_L$ and $\smu_R$ are the SUSY partner of left- and right-handed muon lepton respectively, and the cross section of the former is times larger than the latter. 
Thus $\smu_L$ is more constrained than $\smu_R$, and we mainly study $\smu_L$, denoting it as muon or $\smu$ in this work. 
Since in a real model NMSSM, we calculate the real branching ratios including smuon decay to muon lepton plus the lightest neutralino, that is $Br(\smu^- \to \nino_1 + \mu^-)$. 
We constrain our samples by requiring the cross section timing branching ratio below the corresponding exclusion curves from the ATLAS and CMS collaborations. 

Finally, we got the surviving samples satisfied with all the above constraints, present them in the following figures and carry out corresponding discussions.

\begin{figure*}[htbp]
	\centering
	\includegraphics[width=1\textwidth]{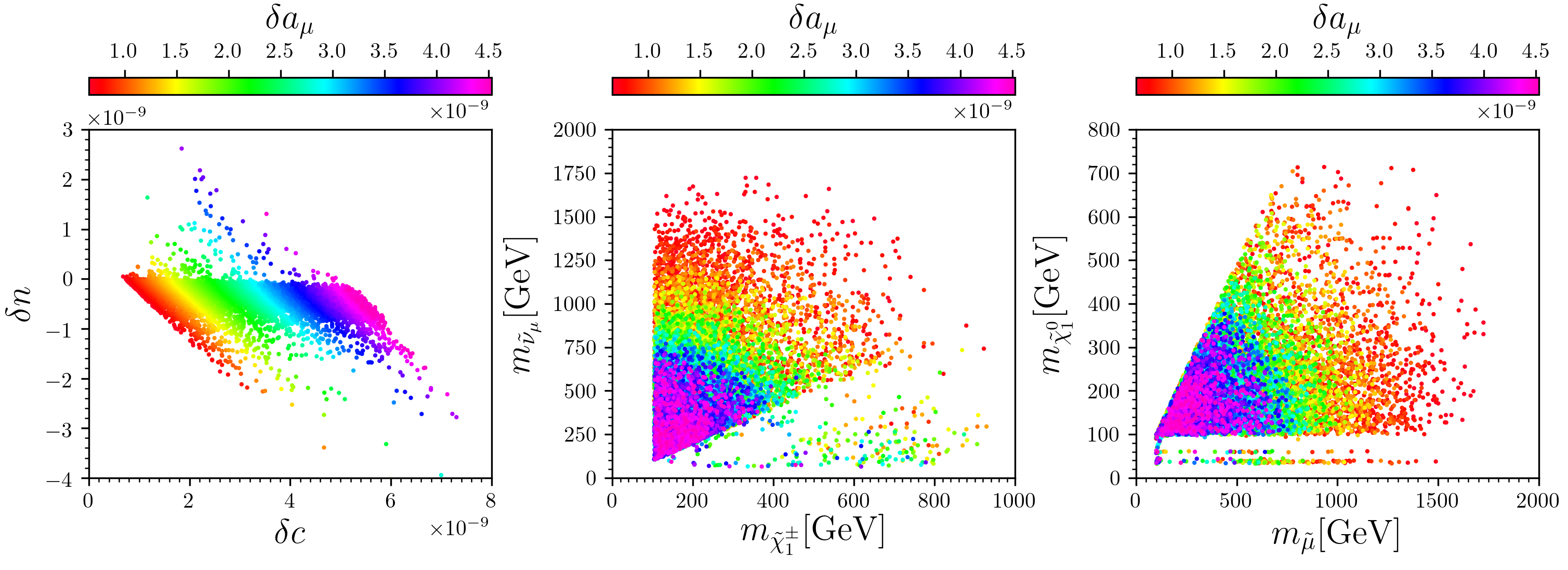}\qquad\qquad
	\vspace{-0.4cm}
\caption{The surviving samples projected in the $\dltan$ versus $\dltac$  (left), $m_{\smu_\mu}$ versus $m_{\cino_1}$ (middle), $m_{\nino_1}$ versus $m_{\smu}$ (right) planes, where $\smu$ is the SUSY partner of the left-handed muon lepton, $\dltamu$ is the SUSY contribution to $\mugmt$, $\dltac$ and $\dltan$ are the chargino-sneutrino and neutralino-smuon contributions to $\dltamu$ respectively. 
In all the planes colors indicate $\delta a_\mu$, and samples with larger  $\delta a_\mu$ are projected on top of smaller ones. 
	}
	\label{fig1}
\end{figure*}

To study the dependence of $\mugmt$ on smuon mass, we display the surviving samples in Fig. \ref{fig1}, from which we can find the following facts:  
\begin{itemize}
\item The $\mugmt$ anomaly can be interpreted within $1\sigma$. 
	For the $1 \sigma$ samples, the muon sneutrinos and left-hand smuons are lighter than $1000\GeV$, the light charginos are lighter than $900\GeV$, and the lightest neutralinos are lighter than $600\GeV$. 
\item 
	For most surviving samples $m_{\cino_1} \approx m_{\nino_1}$ and $m_{\smu} \approx m_{\snu_\mu}$, thus their distributions in the middle and right planes are nearly in the same shapes. 
	We checked that for most of these samples the lightest neutralinos is wino- or higgsino-dominated, same as the light chargino. 
	For the other samples, $m_{\cino_1} > m_{\nino_1}$, and the lightest neutralinos are bino- or singlino-dominated. 
\item 
	The SUSY contribution to $\mugmt$ approximately equal to the sum of chargino-sneutrino and neutralino-smuon loops ($\dltamu \!\approx\! \dltac \!+\! \dltan$), and for most samples the former one is dominated ($\dltac\!>\!\dltan$), even if when the lightest neutralinos and smuon are very light. 
	From Eqs. (\ref{eq:deltan}) and Eq. (\ref{eq:deltac}) we know that make sense, since $g_2^2/32 \gg (g_1^2 \!-\! g_2^2)/192$. 
	Besides we also checked that the contributions are both proportional to $\tanb$, and that $\dltan$ can be both positive and negative because of the different masses of five neutralinos and two smuons. 
\end{itemize}

\begin{figure*}[!htbp]
	\centering
	\includegraphics[width=1\textwidth]{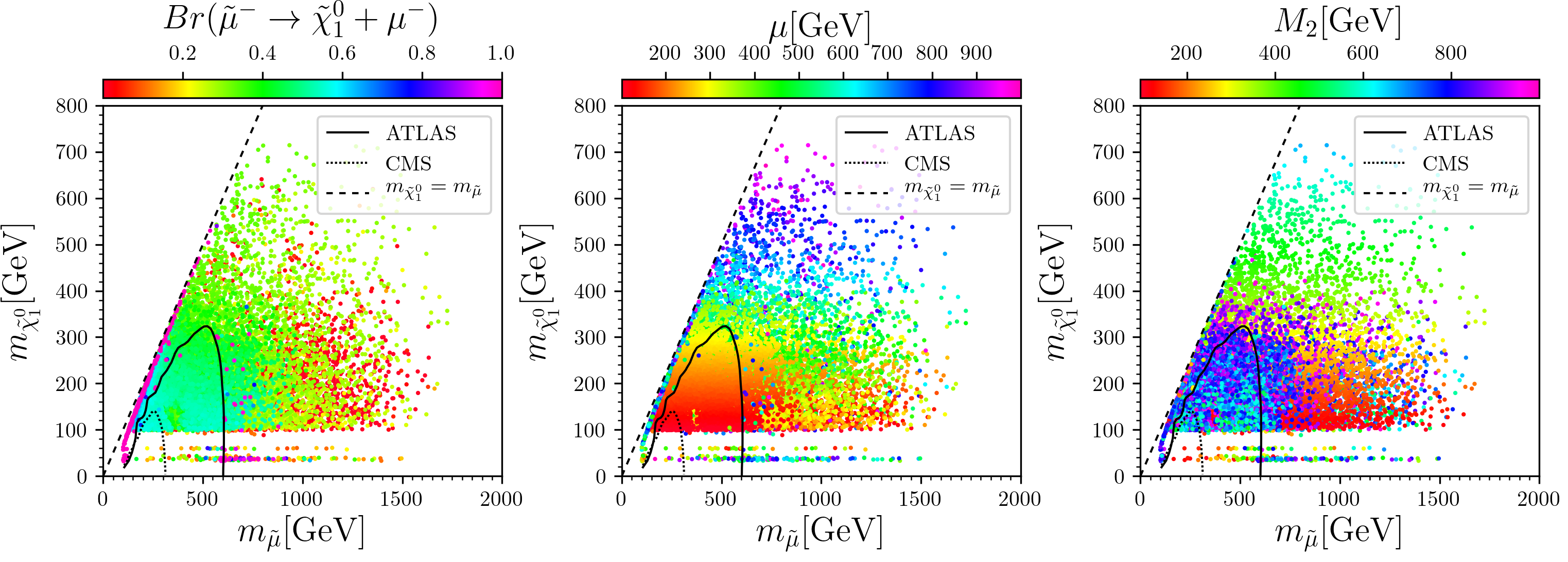}\qquad\qquad
	\vspace{-0.4cm}
	\caption{The surviving samples projected in the $m_{\nino_1}$ versus $m_{\smu}$ planes, where $\smu$ is the SUSY partner of the left-handed muon lepton. 
	Colors indicate the branching ratios $Br(\tilde{\mu}^- \to \tilde{\chi}^0_1 + \mu^-)$ (left), the $\mu$ (middle) and $M_2$ parameters (right) respectively. 
	The dash line of $m_{\nino_1}\!=\! m_{\smu}$ indicates that smuon and $\nino_1$ are mass-degenerate, the solid and dotted curves are the exclusion limits at $95\%$ CL from direct searches for smuons given by the ATLAS \cite{ATLAS:2019lff} and CMS collaborations \cite{CMS:2018eqb} respectively. 
	Samples with larger $Br(\smu^- \to \nino_1 + \mu^-)$ are projected on top of smaller ones. 
	}
	\label{fig2}
\end{figure*}

\begin{figure*}[!htbp]
	\centering
	\includegraphics[width=1\textwidth]{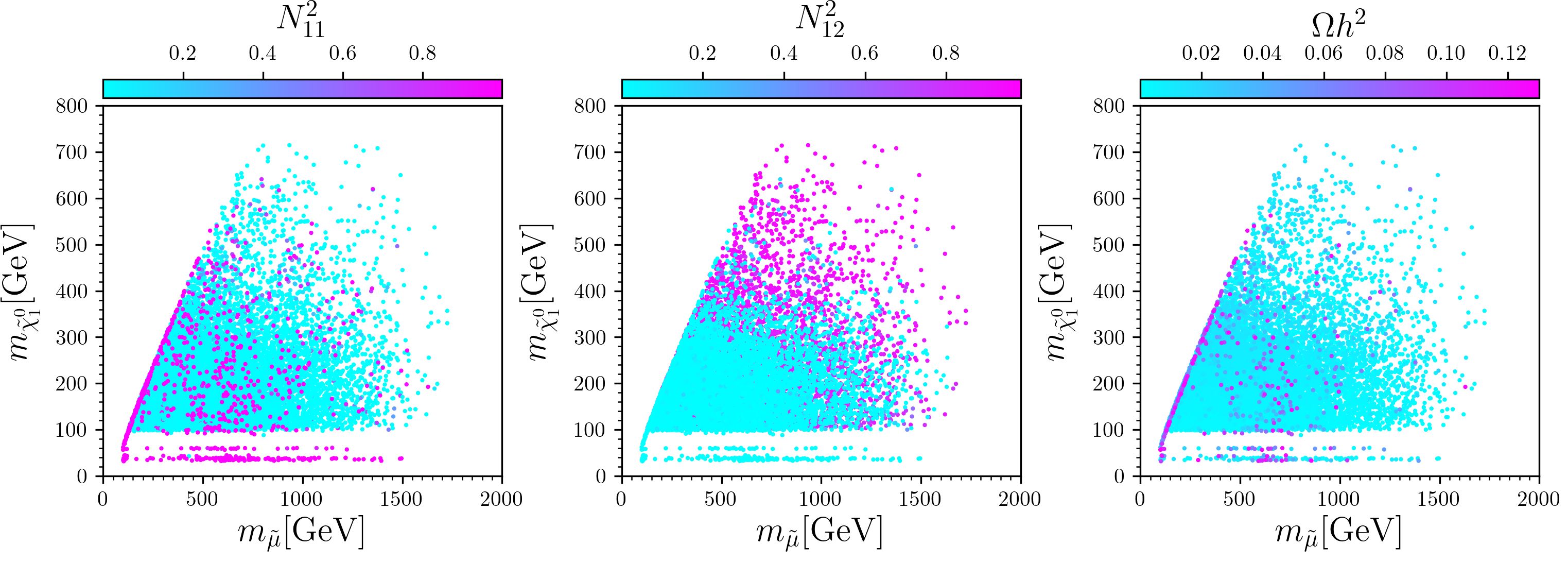}\qquad\qquad
	\vspace{-0.4cm}
	\caption{The surviving samples projected in the $m_{\nino_1}$ versus $m_{\smu}$ planes, where $\smu$ is the SUSY partner of the left-handed muon lepton. 
	Colors indicate $N_{11}^2$ (left), $N_{12}^2$ (middle), $\Omega h^2$ (right), with $N_{11}^2$ and $N_{12}^2$ indicate ratios of bino and wino component in $\nino_1$ respectively,  $\Omega h^2$ indicates the relic density of the LSP dark matter $\nino_1$. 
	Samples with larger $\Omega h^2$ are projected on top of smaller ones. 
	}
	\label{fig3}
\end{figure*}

To study the constraints of smuon searches at the LHC, we display the surviving samples in Fig. \ref{fig2}, where the solid and dotted curves denote the corresponding exclusion limits at 95\% CL by the ATLAS \cite{ATLAS:2019lff} and CMS collaborations \cite{CMS:2018eqb}. 
From this figure we can find that: 
\begin{itemize}
\item For surviving samples below the exclusion curve by ATLAS, all the branching ratios of $Br(\smu^- \!\to\! \nino_1 \!+\! \mu^-)$ are lower than $50\%$, because the large ones have been excluded by the experiments. 
	We checked that for these surviving samples the lightest neutralinos are either higgsino-dominated, with mass approximating to $\mu$ parameter, or wino-dominated, with mass approximating to $M_2$ parameter. 
	Thus $m_{\cino_1} \!\approx\! m_{\nino_1} \!\approx\! \min\{\mu, M_2\}$, and the smuon also decay in a sizable branching ratio to the light chargino plus muon sneutrino. 
\item For surviving samples with mass-degenerate smuon and $\nino_1$, all the branching ratios $Br(\smu^- \!\to\! \nino_1 \!+\! \mu^-)$ are nearly $100\%$, because the current experimental results of direct searches are powerless in the compressed SUSY spectrum. 
\end{itemize}

To illustrate the implication of light smuon on dark matter, we display the surviving samples in Fig. \ref{fig3}, where $N_{11}^2$ denote the bino component in the LSP dark matter $\nino_1$, and $N_{12}^2$ denotes the corresponding wino component respectively. 
From this figure we can learn that:  
\begin{itemize}
\item 
	For most of the surviving samples with the right relic density, or $Z$- or $h_1$-funnel annihilation, the LSP dark matter $\nino_1$ are bino-dominated, some rest are singlino-dominated. 
\item 
	For most of the surviving samples with mass-degenerate smuon and $\nino_1$, the LSP dark matter $\nino_1$ are bino-dominated, and we checked that the dominated annihilating mechanism is slepton annihilations. 
	These samples make up a large part of these with the right relic density only by the LSP $\nino_1$. 
\item 
	For all the surviving samples with wino-dominated $\nino_1$, the relic density is insufficient because of the mass-degenerate charginos, while is also larger than these of higgsino-dominated-$\nino_1$ samples (In all three planes samples with larger $\Omega h^2$ are projected on top of smaller ones, and in the middle plane the wino-dominated-$\nino_1$ samples are covered by the higgsino-dominated-$\nino_1$ ones). 
\end{itemize}

In addition, we also list the detailed information of six benchmark points for further study in Table \ref{tab:benchmark}. 

\begin{table}[!htbp]
\label{tab:benchmark}
\caption{The detail information of six benchmark points, where $N_{i1}^2$, $N_{i2}^2$, $N_{i3}^2\!+\! N_{i4}^2$, and $N_{i5}^2$ indicate ratios of bino, wino, higgsino, and singlino component in neutralinos $\nino_i$ respectively. 
While $\smu_L$ ($\smu_R$) is the SUSY partner of left-handed (right-handed) muon lepton, and $\smu_L$ is the smuon we study in this work.}
\begin{tabular}{ccccccc}
\hline\hline
& ~~~~~~BP1~~~~~~ & ~~~~~~BP2~~~~~~ & ~~~~~~BP3~~~~~~ & ~~~~~~BP4~~~~~~ & ~~~~~~BP5~~~~~~ & ~~~~~~BP6~~~~~~  
\\ \hline $m_{\nino_1} [\GeV]$ & 461 & 34 & 59 & 115 & 336 & 196 
\\ $m_{\nino_2} [\GeV]$ & 712 & 463 & 449 & 150 & 361 & 489 
\\ $m_{\cino_1} [\GeV]$ & 712 & 463 & 449 & 151 & 361 & 488 
\\ $m_{\cino_2} [\GeV]$ & 846 & 610 & 1010 & 346 & 727 & 957 
\\ $N_{11}^2$ & 0.99 & 0.99 & 0.99 & 0.00 & 0.98 & 0.99  
\\ $N_{12}^2$ & 0.00 & 0.00 & 0.00 & 0.00 & 0.01 & 0.00  
\\ $N_{13}^2\!+\! N_{14}^2$ & 0.01 & 0.01 & 0.01 & 0.00 & 0.01 & 0.01  
\\ $N_{15}^2$ & 0.00 & 0.00 & 0.00 & 1.00 & 0.00 & 0.00  
\\ $N_{21}^2$ & 0.00 & 0.00 & 0.01 & 0.00 & 0.02 & 0.01  
\\ $N_{22}^2$ & 0.76 & 0.81 & 0.01 & 0.87 & 0.96 & 0.02  
\\ $N_{23}^2\!+\! N_{24}^2$ & 0.25 & 0.20 & 0.99 & 0.13 & 0.03 & 0.97  
\\ $m_{\smu_L} [\GeV]$ & 471 & 709 & 119 & 143 & 638 & 217 
\\ $m_{\smu_R} [\GeV]$ & 640 & 1480 & 1250 & 368 & 2160 & 575 
\\ $m_{\snu_\mu} [\GeV]$ & 464 & 705 & 92 & 121 & 633 & 203 
\\ $Br(\smu^-\!\to\!\nino_1\!+\! \mu^-)$ & 1.00 & 0.28 & 1.00 & 1.00 & 0.07 & 1.00  
\\ $Br(\smu^-\!\to\!\nino_2\!+\! \mu^-)$ & 0.00 & 0.24 & 0.00 & 0.00 & 0.34 & 0.00  
\\ $Br(\smu^-\!\to\!\cino_1\!+\!\nu_\mu)$ & 0.00 & 0.44 & 0.00 & 0.00 & 0.59 & 0.00 
\\ $\dltamu ~[\times 10^{-10}]$ & 21.8 & 22.8 & 28.2 & 41.7 & 20.5 & 24.3  
\\ $\Omega h^2$ & 0.126 & 0.120 & 0.115 & 0.120 & 0.109 & 0.117  
\\ \hline
\end{tabular}
\end{table}

\section{Conclusions}
\label{sec:summary}

In this work, motivated by the recent supersymmetry (SUSY) search results which prefer most SUSY particles heavy, and $\mugmt$ anomaly which prefers colorless SUSY particles light, we explore the status of a smuon (the SUSY partner of left-handed muon lepton) in the Next-to-Minimal Supersymmetric Standard Model (NMSSM). 
Assuming colored SUSY particles be heavy, and considering numerous experimental constraints including $\mugmt$, SUSY searches, and dark matter, we scan the parameter space in the NMSSM with  $\mathbb{Z}_3$-symmetry, checking the status of colorless SUSY particles and their possible mass order, and paying special attention to the status of smuon confronted with $\mugmt$ and direct SUSY searches. 
We also investigate their implication on dark matter. 

Finally, we draw the following conclusions regarding the smuon (SUSY partner of left-handed muon lepton) and $\mugmt$, SUSY searches, and dark matter in the NMSSM: 
\begin{itemize}
\item 
	The dominated SUSY contributions to $\mugmt$ come from the chargino-sneutrino loops, but the $\mugmt$ anomaly can also constrain the mass of smuon seriously because of the mass-degenerate relation between smuon and muon sneutrino. 
	To interpret the $\mugmt$ anomaly at $1\sigma$ ($2\sigma$) level, the smuon need to be lighter than $1\TeV$ ($1.8\TeV$). 
\item 
	when $\nino_1$ is wino- or higgsino-dominated, the smuon and muon sneutrino, the $\cino_1$ and $\nino$, are usually mass-degenerate respectively. 
	Thus the branching ratio $Br(\smu^- \!\to\! \nino_1 \!+\! \mu^-)$ is suppressed to lower than $50\%$, and smuon can escape from the direct searches with low mass, e.g., $300\GeV$. 
\item 
	When smuon and $\nino_1$ are mass-degenerate, the smuon can be as light as $200\GeV$, while the $\nino_1$ are usually bino-dominated, or some singlino-dominated, and its relic density can most likely reach the observed value, and the dominated annihilating mechanism is slepton annihilation. 
\end{itemize}

\begin{acknowledgments}
\paragraph*{Acknowledgements:}
	This work was supported by the National Natural Science Foundation of China (NNSFC) under grant Nos. 11605123, and the Startup Research Foundations of Henan University. 
\end{acknowledgments}

\bibliography{ref}

\end{document}